\documentclass[pra,twocolumn,a4paper,showpacs,aps,10pt]{revtex4-2}
\usepackage{physics} 
\usepackage{amsmath}
\usepackage{amsthm}
\usepackage{amssymb} 
\usepackage{comment}
\usepackage{graphicx}
\usepackage{caption}
\usepackage{subcaption}
\usepackage{quantikz}
\usepackage{tikz}
\usepackage{dsfont}
\usepackage{bbold}

\newtheorem{definition}{Definition}

\newtheorem{theorem}{Theorem}
\newtheorem{corollary}{Corollary}
\newtheorem{lemma}{Lemma}

\newcommand{\Pbar}{\overline{\mathcal{P}}}
\newcommand{\Sb}{\overline{S}}
\newcommand{\bs}[1]{\ensuremath{\boldsymbol{#1}}}

\newcommand{\mcB}{\mathcal{B}}
\newcommand{\Pgrpd}{\mathcal{P}_d^n}
\newcommand{\clif}{{\rm Cl}}

\begin{document}

\date{September 2026}

\title{\bf Causal inequalities witness non-stabilizerness}
\author{Leonardo Vaglini}
\affiliation{Aix-Marseille University, CNRS, LIS, Marseille, France}
\author{Nasra Daher Ahmed}
\affiliation{Aix-Marseille University, CNRS, LIS, Marseille, France}
\author{Ravi Kunjwal}
\affiliation{Aix-Marseille University, CNRS, LIS, Marseille, France}

\begin{abstract}
Stabilizer operations describe a fragment of quantum theory that is known to be efficiently classically simulable, thanks to the Gottesman-Knill theorem. For this reason, nonstabilizer resources such as magic states are necessary for universal quantum computation. Interestingly, the operational and axiomatic approaches to the resource theory of magic differ: the set of free operations in the former, namely, stabilizer operations (SO), is strictly smaller than that in the latter, namely, completely stabilizer preserving operations (CSPO). A simple example showing the separation is given by a three-qubit stabilizer product basis whose states cannot be perfectly discriminated using SO, but which do admit perfect discrimination using CSPO. Such an ensemble of states is said to exhibit \emph{nonstabilizerness without magic} (NSWM). Here we obtain a principled understanding of this phenomenon, proving necessary and sufficient conditions for its existence. We first derive a simple criterion to decide whether, given a stabilizer basis, its states can be perfectly discriminated using stabilizer operations alone. We then consider the case where the stabilizer basis contains only product states and use its link with process functions---classical models of paradox-free causal loops---to prove the following: the states in a stabilizer product basis require nonstabilizerness for perfect discrimination if and only if the corresponding process function violates a causal inequality. This provides a new operational meaning to causal inequality violations as witnesses of nonstabilizerness, a form of computational nonclassicality.

\end{abstract}

\maketitle

\section{Introduction}

The Gottesman-Knill theorem \cite{gottesman1997stabilizercodesquantumerror,gottesman1998heisenbergrepresentationquantumcomputers,PhysRevA.70.052328,articleNest} places a fundamental limit on the power of quantum computation that must be overcome to achieve a quantum speedup over classical computers. It establishes that any quantum computation performed using stabilizer operations (SO), \textit{i.e.}, operations built out of Clifford gates, preparation of stabilizer states, and Pauli measurements performed adaptively, can be efficiently classically simulated. Therefore, a quantum computational advantage requires the use of non-Clifford gates or the preparation of nonstabilizer states known as magic states. 

In the operational approach to the resource theory of magic, stabilizer operations define the set of free operations \cite{RevModPhys.91.025001,Veitch_2014}. In the axiomatic approach \cite{10.1098/rspa.2019.0251,PRXQuantum.2.010345} to such a resource theory, the free operations are completely stabilizer preserving operations (CSPO), \textit{i.e.}, channels which preserve the stabilizer polytope, even in the presence of an ancilla. These two different approaches have been shown to be inequivalent, in the sense that $\rm SO\subsetneq CSPO$ \cite{10.1063/5.0085774}. An operational instance of this separation is provided in Ref.~\cite{kwon2025nonstabilizernessmagicclassicallysimulatable}, showing that the SHIFT ensemble
\begin{equation}
\begin{aligned}
\mathcal{B}_{\rm SHIFT}=\{&\ket{000},\ket{+01},\ket{01+},\ket{01-},\\
&\ket{1+0},\ket{-01},\ket{1-0},\ket{111}\}
\end{aligned}
\end{equation}
cannot be discriminated via stabilizer operations alone, even though such states can be easily prepared using stabilizer operations,
thus exhibiting \emph{nonstabilizerness without magic} (NSWM) \cite{kwon2025nonstabilizernessmagicclassicallysimulatable}. 

Curiously, the SHIFT ensemble also demonstrates another interesting separation, namely, between the operational resource theory of entanglement based on local operations and classical communication (LOCC) and the axiomatic resource theory of entanglement based on separable operations (SEP), \textit{i.e.}, $\rm LOCC \subsetneq SEP$. This is known as quantum nonlocality without entanglement (QNLWE) \cite{PhysRevA.59.1070}. Recent work \cite{PhysRevLett.131.120201,dourdent2025paradoxfreeclassicalnoncausalityunambiguous} has shown an intimate link between the phenomenon of QNLWE and process functions \cite{Baumeler_2016}, the latter being classical models of paradox-free causal loops originating from the process-matrix framework \cite{Oreshkov2012aa}. Specifically, 
allowing the parties to communicate via the AF/BW process function \cite{Baumeler_2016,PhysRevLett.131.120201}, \textit{i.e.},
\begin{equation}
    x_1=a_3(a_2\oplus 1),\quad x_2=a_1(a_3\oplus 1), \quad x_3=a_2(a_1\oplus 1),
\end{equation}
allows them to implement the separable measurement that perfectly discriminates the SHIFT basis \cite{PhysRevLett.131.120201} using local operations alone. Here $x_i$ and $a_i$ are, respectively, the input and output variables of party $i$ and the process function describes how the input of each party is determined by the outputs of the others. 

Given that the SHIFT basis proves both separations ($\rm LOCC \subsetneq SEP$ and $\rm SO \subsetneq CSPO$), it is natural to wonder if the separation $\rm SO \subsetneq CSPO$ might also admit a causal reading. An immediate obstacle is that stabilizer operations are fundamentally different from LOCC: SO allows entangled measurements (forbidden in LOCC) and LOCC allows nonstabilizer measurements (forbidden in SO). This means that the methods used in the QNLWE case do not translate to the NSWM case.

We overcome this obstacle and prove two main results for multiqudit stabilizer bases with qudits of prime dimension: firstly, a necessary and sufficient criterion for a stabilizer basis to exhibit NSWM (Theorem \ref{thm:criterion}), and secondly, that a stabilizer product basis displays NSWM if and only if the associated process function is noncausal (Theorem \ref{thm:main}). 

Our results provide a new operational meaning to causal inequality violations as witnesses of nonstabilizerness, \textit{i.e.}, a form of computational nonclassicality. This opens up several new areas of inquiry at the intersection of these two fields, with strong potential for a fruitful interaction between them.




\section{State discrimination in the stabilizer subtheory.}

We begin by introducing the Pauli group and the stabilizer formalism in dimension $d$ \cite{PhysRevA.71.042315,Tolar_2018,10.1063/1.2393152}, assuming $d$ prime throughout the rest of the work.
Let $\omega=e^{\frac{2\pi i}{d}}$ for odd prime $d$ and $\omega=i$ for $d=2$. The Pauli group is defined in terms of the shift and boost operators $X$ and $Z$ respectively. Their action on the computational basis is given by
\begin{equation}
    \begin{aligned}
        X\ket{a}&=\ket{a\oplus 1} \\
        Z\ket{a}&=\zeta^a\ket{a}
    \end{aligned}
\end{equation}
where $\zeta=e^{\frac{2\pi i}{d}}$. Both $Z$ and $X$ are order $d$, meaning that $Z^d=X^d=\mathds{1}$, and satisfy the following commutation relation $XZ=\zeta^{-1} ZX$. For odd prime $d$, the Pauli group is generated by $X$ and $Z$ as follows
\begin{equation}\label{eq:PauliGroup}
    \mathcal{P}_d=\{\omega^cX^aZ^b|a,b,c\in\mathbb{Z}_d\}.
\end{equation}
Every nontrivial $P\in\mathcal{P}_d$ has order $d$. This implies that each $\omega^c\mathds{1}\neq P\in\mathcal{P}_d$ has $d$ different nondegenerate eigenvalues $\{ \zeta^m \}_{m\in \mathbb{Z}_d}$.  
The $n$-qudit Pauli group $\mathcal{P}_d^{n}$ is generated by the tensor product of single-system Pauli operators. Each $P\in\mathcal{P}_d^n$ is of the following form
\begin{equation}\label{eq:tensorPauli}
P=\omega^cX^{a_1}Z^{b_1}\otimes\dots\otimes X^{a_n}Z^{b_n},\quad c,a_i,b_i\in\mathbb{Z}_d.
\end{equation}
For $d=2$, the Pauli group is given by equations \eqref{eq:PauliGroup} and \eqref{eq:tensorPauli} provided that $c\in\{0,1,2,3\}$.
    We then consider the projective Pauli group $\Pbar_d^n=\mathcal{P}_d^n/\langle\omega\mathds{1}\rangle$, i.e., the Pauli group quotiented by the phase group. As representative of a given class in $\Pbar_d^n$ we take, for odd prime $d$, the
    operator with coefficient 1, and for $d=2$ the operator $P_1\otimes\dots\otimes P_n$ with $P_k\in\{\mathds{1},X,Y,Z\}$, where $Y:=iXZ$. In both cases $P^d=\mathds{1}$, so the spectrum of $P$ is contained in $\{\zeta^m\}_{m\in\mathbb{Z}_d}$ and $\Pi^P_m=\frac{1}{d}\sum_t\zeta^{-mt}P^t$ is the projector onto its $\zeta^m$-eigenspace. When $P\in\Sb_\phi$, by $P\ket\phi$ we always mean the result of applying a representative of the equivalence class $P$ to the vector $\ket{\phi}$.
    As a notational convention, we denote a local operator $\mathds{1}\otimes P_\ell\otimes\mathds{1}$ simply as $P_\ell$. Given a stabilizer state $\ket{\phi}$, we denote by $\mathcal{S}_\phi=\{P\in\Pgrpd:P\ket{\phi}=\ket{\phi}\}$ the corresponding stabilizer group, and with $\Sb_\phi=\{P\in\Pbar_d^n:\exists c\in\{0,\dots,d-1\}\text{ s.t. }P\ket{\phi}=\zeta^c\ket{\phi}\}$ 
    the subgroup of $\Pbar_d^n$ obtained by applying the quotient map $\pi:\Pgrpd\rightarrow\Pbar_d^n$ to the stabilizer group $S_\phi$. The Clifford group is the normalizer of the Pauli group, i.e., it is the subgroup of unitary operations defined as $\clif_d^n=\{C:CPC^\dagger\in\Pgrpd,\;\forall P\in\Pgrpd\}$.

An adaptive stabilizer discrimination protocol consists of: i) measurements of Pauli operators, ii) Clifford operations, and iii) the preparation of an arbitrary number of ancillary qudits in the state $\ket{0}$. Operations i) and ii) can be performed adaptively, depending on the outcomes of previous measurements. The superoperator associated with such a protocol is given by
\begin{equation}
    \mathcal{D}(\cdot)=\sum_{\bs{m}}C_k\Pi_{m_k}^{P_k}\dots  C_1\Pi_{m_1}^{P_1} \cdot \Pi^{P_1}_{m_1}C_1^\dagger\dots \Pi^{P_k}_{m_k}C_k^\dagger
\end{equation}
\noindent
where each term $C_\ell\Pi^{P_\ell}_{m_\ell}$ generally depends on the previous $\ell-1$ outcomes $m_1,\dots,m_{\ell-1}$. Since the $C$ operations are Clifford and each $\Pi^{P}_m$ is a Pauli measurement, we have $C\Pi_m^PC^\dagger=\Pi^{P'}_m$ for a suitable Pauli operator $P'$. Therefore, all the Clifford unitaries can be moved to the end, and the protocol is essentially given by a sequence of Pauli measurements
\begin{equation}\label{eq:KrausProt}
\Pi_{m_k}^{P_k}\circ\dots\circ\Pi_{m_1}^{P_1}
\end{equation}
with $k$ arbitrarily large. Therefore, we can always think of stabilizer discrimination protocols as adaptive sequences of stabilizer measurements, each described by a Pauli operator. Note that the operators in equation \eqref{eq:KrausProt} can generally act on both the system and ancillary qudits. However, using ancillas turns out to provide no advantage, as a consequence of the following lemma.

\begin{lemma}\label{lem:ancilla_removal}
    Consider $\ket{\psi}\ket{0}^{\otimes k}$ with $\ket{\psi}$ an arbitrary stabilizer state on $n$ system qudits ($d$ being prime). Let $P$ be a Pauli measurement that does not commute with at least one of the generators of the ancilla. Let $\Pi_m^P$ be the projector corresponding to the outcome $m$.Then the probability distribution of the outcomes is $p(m)=1/d$ for every $m\in\{0,\dots,d-1\}$. Moreover, there exists a Clifford operation $U(P,m)\in\clif_d^{n+k}$ independent of $\ket{\psi}$ such that:
    \begin{equation}
    \begin{aligned}
        U(P,m)\frac{\Pi^P_m}{\sqrt{p(m)}}\ket{\psi}\ket{0}^{\otimes k}=\ket{\psi}\ket{0}^{\otimes k},
    \end{aligned}
    \end{equation} 
    namely, the original state $\ket{\psi}\ket{0}^{\otimes k}$ is restored.
\end{lemma}

The lemma shows that any measurement that does not commute with the generators of the ancillary systems does not provide any information about the state at hand and can be inverted by means of a suitable Clifford operation that solely depends on the measurement outcome. Therefore, in the discrimination problem of sets of stabilizer states, it is sufficient to consider the ancilla-free case, not only for $d=2$ (as already shown in \cite{kwon2025nonstabilizernessmagicclassicallysimulatable}) but for every prime dimension $d$. As we have characterized the protocols, we formally define what it means for a stabilizer ensemble to exhibit nonstabilizerness without magic. 

\begin{definition}[\cite{kwon2025nonstabilizernessmagicclassicallysimulatable}]
    A set of stabilizer states $\mathcal{E}$ is said to exhibit nonstabilizerness without magic if it cannot be perfectly discriminated through an adaptive stabilizer discrimination protocol. 
\end{definition}

As shown in Ref.~\cite{kwon2025nonstabilizernessmagicclassicallysimulatable}, not all stabilizer state ensembles can be perfectly discriminated using SO. This has been proved for the set $\tilde\mcB_{\rm SHIFT}$, the subset of the SHIFT basis $\mcB_{\rm SHIFT}$ obtained by removing the states $\ket{000}$ and $\ket{111}$. 
Note that $\tilde\mcB_{\rm SHIFT}$ has the following property: for each qubit there are at least two different local bases, the $Z$ and the $X$ eigenbases. 

In the general case of $n$-qudit systems, we can formalize the above observation as follows. Given an orthonormal stabilizer set $\mcB=\{\ket{\phi_i}\}_{i=1}^{|\mcB|}$ on $n$ qudits, we define the subgroup $\Sb_\mcB=\bigcap_i \Sb_{\phi_i}$. When $\mcB$ is a product set, this subgroup retains information about which qudits have a constant local basis throughout all states in $\mcB$, and in particular one has $\Sb_\mcB=\{\mathds{1}\}$ if and only if no qudit has a constant basis. Indeed, for  the $\tilde\mcB_{\rm SHIFT}$ we have $\Sb_{\tilde\mcB_{\rm SHIFT}}=\{\mathds{1}\}$. 

What does this subgroup tell us about the perfect discriminability of a stabilizer ensemble in general?

In general, the condition $\Sb_\mcB=\{\mathds{1}\}$ is not sufficient to have NSWM, since removing a state from $\tilde\mcB_{\rm SHIFT}$ leads to a perfectly distinguishable ensemble \cite{kwon2025nonstabilizernessmagicclassicallysimulatable}, while it becomes sufficient for bases, as we will show. On the other hand, for bases it is not necessary: indeed, consider the basis $\mcB$ given by the union of $\ket{0}\otimes\mcB_{\rm SHIFT}$ and $\ket{1}\otimes \mcB_{\rm comp}$, where $\mcB_{\rm comp}$ is the computational basis of three qubits. For the first qubit, there is only the $Z$ eigenbasis. However, once it has been measured, the obstruction to perfect discrimination in this example arises from $\mcB_{\rm SHIFT}$, which is the ensemble to be discriminated in the branch corresponding to outcome $m=0$. Thus, the obstruction comes again from the $\Sb_\mcB=\{\mathds{1}\}$, but applied to one of the branches after the first measurement. 

We show that this property becomes a sufficient recursive condition when the ensemble is a basis, thereby capturing cases where only a subset of systems may be such that none of them have a constant local basis. Furthermore, the condition is not tied to stabilizer product bases: it works for any stabilizer basis, including those with entangled states where talk of a ``constant local basis" for each system has no meaning.

\begin{lemma}\label{lem:auxlem_2}
Let $\mcB=\{\ket{\phi_i}\}_{i=1}^{|\mcB|}$ be an orthonormal stabilizer basis, $P\in \Pbar_d^n$ and $\Pi_m^P$ its projection corresponding to outcome $m$. Then the following are equivalent:
\begin{enumerate}
    \item \label{it:pmstate_ort} $\langle\phi_i|\Pi_m^P|\phi_j\rangle=0$ for all $m\in\{0,1,\dots,d-1\}$ and $i\neq j$;
    \item $\langle\phi_i|P^t|\phi_j\rangle=0$ for all $t\in\{0,1,\dots,d-1\}$ and $i\neq j$;
    \item $P\in\Sb_\mcB$.
\end{enumerate}
In particular, every $P\not\in\Sb_\mcB$ breaks the orthogonality of at least one pair of states.
\end{lemma}
In a sense, $\Sb_\mcB$ contains all the deterministic measurements. The above lemma essentially says that when $\mcB$ is a basis, performing any nondeterministic measurements immediately causes two post-measurement states to be nonorthogonal. Note that for any product ensemble $\mcB$, a nonlocal Pauli $P\in\Sb_\mcB$ if and only if each local factor $P_\ell\in\Sb_\mcB$. 
This implies that in order to assess whether an SPB is perfectly discriminable it is enough to check local Pauli measurements only. 


Before presenting our first theorem we state the following standard lemma, which also sets up the notation.
\begin{lemma}\label{lem:auxlem3}
    Let $\mcB$ be an orthonormal stabilizer basis, suppose $\Sb_\mcB\neq\{\mathds{1}\}$ and consider the representatives of $r$ independent generators of $\Sb_\mcB$, and let $V_{\chi}$ their joint eigenspaces with eigenvalues $\chi\in\{\zeta^c\}_{c\in\mathbb{Z}_d}^r$. Then the following holds:
    \begin{enumerate}
    \item\label{it:auxlem3_1} $\dim V_\chi=d^{n-r}$ and $|\mcB\cap V_\chi|=d^{n-r}$, therefore $\mcB\cap V_\chi$ is a basis of $V_\chi$. 
    \item\label{it:auxlem3_2} For every outcome $\chi$, there exists a Clifford $U_\chi\in\clif_d^n$ such that $U_\chi(\mathcal{B}\cap V_\chi)= \mathcal{B}_\chi\otimes\ket{0}^{\otimes r}$ for some stabilizer basis $\mathcal{B}_\chi$ of $(\mathbb{C}^d)^{\otimes(n-r)}$.
    \end{enumerate}
\end{lemma}
This lemma says that once we have measured all the operators yielding a deterministic result, the discrimination problem is reduced to the discrimination of the bases $\mcB_\chi$, one for every joint eigenvalue $\chi$. We can now state and prove the following criterion:

\begin{theorem}\label{thm:criterion}
Let $\mcB$ be a stabilizer basis on $n$ qudits with $n\geq2$. Then $\mcB$ is  perfectly discriminable if and only if $\Sb_\mcB\neq\{\mathds{1}\}$ and $\mathcal{B}_\chi$ is perfectly discriminable for every joint eigenvalue $\chi$ of the independent generators of $\Sb_\mcB$.
\end{theorem}

This represents a recursive criterion that allows us to determine whether a given stabilizer basis $\mcB$ is perfectly discriminable by looking at the sets $\Sb_{\mcB_\chi}$ at each step. As a corollary of this theorem, we have the following:

\begin{corollary}\label{cor:NSWM}
    Let $\mcB$ be a stabilizer basis on $n$ qudits with $n\geq 2$. Then, if $\Sb_\mcB=\{\mathds{1}\}$, $\mcB$ exhibits NSWM.
\end{corollary}

The fact that $\mcB_{\rm SHIFT}$ exhibits NSWM is now a consequence of $\Sb_{\mcB_{\rm SHIFT}}=\{\mathds{1}\}$. 

\section{Non-causality meets non-stabilizerness}

We now specialize to product bases. We show that any stabilizer product basis (SPB) $\mcB$ is associated with a unique process function, and furthermore, $\mcB$ exhibits NSWM if and only if the associated process is noncausal.

A process function \cite{Baumeler_2016} is a mathematical object that describes deterministic classical communication among a given set of parties. It is a function $\omega:\times_k\mathcal{A}_k\rightarrow \times_k\mathcal{X}_k$ that maps the collective outputs of the parties to their inputs, ruling who communicates with whom. Here $\mathcal{X}_k$ and $\mathcal{A}_k$ denote the sets of inputs and outputs of party $k$ respectively. Such objects are tightly related to unambiguous orthonormal product bases \cite{dourdent2025paradoxfreeclassicalnoncausalityunambiguous}, of which SPBs represent a subset. 

Indeed, an unambiguous basis $\mcB = \{\ket{\phi_i} = \otimes_{k=1}^n\ket{\phi_i^k}\}_{i =1}^{|\mcB|}$
is a product basis such that, for every party $k$, two local vectors $|\phi_i^k\rangle$, $|\phi_j^k\rangle$ are orthogonal if and only if they belong to the same local basis. SPBs are such that local bases are mutually unbiased, hence they satisfy the latter condition, implying that any SPB is unambiguous. This entails that SPBs can be represented in terms of a suitable process function and a set of local Clifford unitaries $\{U_k^{x_k}\}_{k=1}^n$ as follows \cite{dourdent2025paradoxfreeclassicalnoncausalityunambiguous}
\begin{equation}\label{eq:pf_rep}
\mcB=\{\bigotimes_kU_k^{\omega_k(\boldsymbol a_{\backslash k})}\ket{\boldsymbol a}|\bs{a}\in\bs{\mathcal{A}}\}.
\end{equation}
In this representation, local unitaries $\{U_k^{x_k}\}_{x_k\in\mathcal{X}_k}$ retain information about which local bases appear in $\mcB$, while the process function $\omega$ determines the global structure of the basis. In particular $\omega$ tells us which qudits have more than one basis (\textit{i.e.}, any qudit $k$ for which $\omega_k$ is not a constant function), thus determining its perfect discriminability.

More specifically, what determines the distinguishability of $\mcB$ is the causal structure of $\omega$.
The notion of causality for a process function can be characterized recursively \cite{dourdent2025paradoxfreeclassicalnoncausalityunambiguous}: Firstly, there must exist a party $k$ in the global past of all others, which means that the corresponding component $\omega_k$ is constant. In such a case, the party $k$ always receives a constant input. Secondly, for all such parties and for all their possible outputs, the remaining ones still communicate in a definite causal order. Read in the contrapositive, a noncausal process might admit a partial causal order among some parties, but there will always exists a configuration of such parties such that the remaining ones do not have a global past.

A genuinely noncausal process is such that no party is in the global past of all the others, meaning that none of the $\omega_k$'s are constant.
If this is the case, it is not difficult to realize that the corresponding basis must have $\Sb_\mcB=\{\mathds{1}\}$ and according to Corollary \ref{cor:NSWM} has NSWM. We have the following theorem, whose complete proof is in Appendix:

\begin{theorem}\label{thm:main}
    For every $n\geq 1$, every SPB $\mcB$ is perfectly discriminable with stabilizer operations iff the associated process function is causal.
\end{theorem}

It is known that every noncausal process function violates a causal inequality \cite{Baumeler_2016,dourdent2025paradoxfreeclassicalnoncausalityunambiguous}. Therefore, the above result can also be alternatively phrased as follows:

\begin{corollary}
    For every $n\geq 1$, every SPB $\mcB$ exhibits NSWM iff the associated process function violates a causal inequality. 
\end{corollary}

The above theorem can be read as a trade-off between nonstabilizerness (a type of computational nonclassicality) and causal order: giving up on causal order allows us to implement the  nonstabilizer operation required for perfect discrimination using stabilizer operations alone (\textit{e.g.}, using the protocol outlined in Ref.~\cite{PhysRevLett.131.120201}).

\begin{section}{Discussion}
   In the present work, we have considered the general state discrimination problem for orthonormal complete sets of stabilizer states within the operational resource theory of magic. We have provided a simple recursive criterion formalized in Theorem \ref{thm:criterion} that establishes whether a given stabilizer basis can be perfectly discriminated using stabilizer operations. This is essentially based on the stabilizer subgroup $\Sb_\mcB$ corresponding to those measurements which give a deterministic result on all the states of the basis, and it applies to any kind of stabilizer basis, including those with entangled states.  
    
    Following this, we focused on stabilizer product bases, which naturally admit a process function representation. The tight association between process functions and stabilizer product bases is guaranteed by the fact that such bases are always unambiguous. Then we showed that an SPB is perfectly distinguishable under SO if and only if the associated process function is causal. In other words, an SPB exhibits NSWM if and only if the corresponding process function violates a causal inequality. Theorem \ref{thm:main} precisely identifies the distinction between causal and noncausal process functions as the demarcating boundary between those bases that can be discriminated through SO and those exhibiting NSWM. 
    
    We remark that the class of SO is inherently different from that of LOCC, since the former allows for nonlocal measurements.
    Because of the fact that the set of states to be distinguished in an SPB is separable, and at the same time the set of allowed measurements in SO is rather constrained (even if it allows for nonlocal measurements), one might be tempted to disregard the power of nonlocal measurements for state discrimination. Such a move is legitimate as long as we are only interested in assessing whether the SPB is \textit{perfectly} discriminable. However, this does not imply that general stabilizer protocols can never beat LOCC ones at \textit{approximate} distinguishability when an SPB exhibits NSWM. We leave open the problem of whether a gap between the two success probabilities---LOCC vs.~SO---in the discrimination task exists. In this respect, it would be also interesting to see whether there is a tight relation between the guessing probability of the ``guess the process function game" of Ref.~\cite{dourdent2025paradoxfreeclassicalnoncausalityunambiguous} and the discrimination success probability.

    
    Theorem \ref{thm:main} characterizes NSWM for SPBs. We point out that there also exist stabilizer bases that can contain entangled stabilizer states, such as the one obtained from the completion of the set $\{\ket{01+},\ket{1+0},\ket{+01},\ket{---}\}$. This stabilizer basis still exhibits NSWM (from Theorem \ref{thm:criterion}) and, at the same time, its subset $\{\ket{01+},\ket{1+0},\ket{+01},\ket{---}\}$ can be perfectly discriminated with local Clifford operations and the AF/BW process function \cite{Baumeler_2016}, showing the same trade-off between causal order and nonstabilizerness enjoyed by all NSWM SPBs. It is thus clear that one can obtain NSWM entangled ensembles by applying suitable multi-qudit Clifford gates to SPBs. What is not known is whether there exist NSWM entangled ensembles that are not Clifford-equivalent to a product one; in the affirmative case, any potential interpretation of NSWM in terms of noncausal communication must relax the assumption of logical consistency. It would also be worth studying this question in relation to the existence of unextendible stabilizer bases, \textit{i.e.}, sets of orthonormal stabilizer states that cannot be completed to an orthonormal stabilizer basis \cite{frembs2026unextendiblestabiliserbases}.

    Conceptually, our results show how the phenomenon of NSWM can be revisited from a causal perspective, proving that causal inequality violations can serve as witnesses of nonstabilizerness. Technically, they fully generalize the observation of NSWM in Ref.~\cite{kwon2025nonstabilizernessmagicclassicallysimulatable} to arbitrary prime dimensions and provide a necessary and sufficient characterization for NSWM. As such, they provide a new causality-inspired pathway to understanding the resource aspects of nonstabilizerness.

\begin{acknowledgments}
This work received support from the French government under the France 2030 investment plan, as part
of the Initiative d’Excellence d’Aix-Marseille Universit´eA*MIDEX, AMX-22-CEI-01
\end{acknowledgments}
    
\end{section}

\bibliographystyle{unsrturl}
\bibliography{biblio} 

\appendix
\section{Proof of theorem \ref{thm:criterion}}

\subsection{Proof of lemma \ref{lem:ancilla_removal}}
\label{app:ancilla_removal_proof}

We just consider odd prime $d$, since the $d=2$ case is treated in \cite{kwon2025nonstabilizernessmagicclassicallysimulatable}. Let $\mathcal{S}_{\ket{\psi}\otimes \ket{0}^{\otimes k}}=\langle g_1,\dots,g_n,Z_{n+1},\dots,Z_{n+k}\rangle$ be the stabilizer of the state $\ket{\psi}\otimes\ket{0}^{\otimes k}$. Suppose that the operator $P$ does not commute with a generator of the ancilla, say $Z_{n+h}$, and that the measurement outcome is $m$. We now update the stabilizer in the following way: we replace $Z_{n+h}$ with $\zeta^{-m}P$. The generators of $\mathcal{S}_{\ket{\psi}\otimes \ket{0}^{\otimes k}}$ which commute with $P$ are left unchanged. Finally, consider a generator $g_{\ell}$ which does not commute with $P$, thus satisfying the relation 
\begin{equation}
    Pg_\ell=\zeta^{\phi_\ell}g_\ell P.
\end{equation}
Suppose that the $n+h$-th factor $P_{n+h}$ of $P$ contains a number $x$ of shift operators $X$ and a number $z$ of operators $Z$. Then the fundamental relation $XZ=\zeta^{-1}ZX$ implies
\begin{equation}
    PZ_{n+h}=\zeta^{-x}Z_{n+h} P.
\end{equation}
Therefore, the operator $g Z_{n+h}^q$ (we will omit the subscript $\ell$ to simplify the notation) satisfies the following commutation relation with $P$
\begin{equation}
    PgZ^q=\zeta^{\phi-qx}gZ^qP.
\end{equation}
Thus, the operator $gZ^q$ commutes with $P$ if and only if $\phi=qx\;{\rm (mod\;d)}$, which always has a solution in $q$ for prime $d$. Moreover, we can readily see that this operator stabilizes the post-measurement state, since $gZ^q$ stabilizes $\ket{\psi}\otimes\ket{0}^{\otimes k}$ and commutes in particular with the projectors $\Pi_m^P$
Now consider the following unitary operator
\begin{equation}
    U=\sum_{\ell=0}^{d-1}\left(PX_{n+h}^{-x}\right)^\ell\ket{\ell}\bra{\ell}_{n+h}
\end{equation}
where $\ket{\ell}\bra{\ell}_{n+h}$ acts non-trivially as $\ket{\ell}\bra{\ell}$ only on the $n+h$-th qubit, and is the identity elsewhere. From now on, we will omit the subscript whenever there is no risk of confusion. This is a Clifford operation. It is straightforward to see that $UZU^\dagger=Z$ and that $UgU^\dagger=g$ for every generator $g$ that commutes with $P$. Moreover, its action on the operators $P$ and $gZ^q$ is given by
\begin{equation} \label{eq:quditcliff}
    \begin{aligned}
        UPU^\dagger & =(PX_{n+h}^{-x})^x P Z^{xz}, \\
        UgZ^qU^\dagger & = gZ^{xq}Z^q. \\
    \end{aligned}
\end{equation}
Furthermore, it is not difficult to see that $UPX_{n+h}^{-x}U^\dagger=PX_{n+h}^{-x}$, namely, $PX_{n+h}^{-x}$ is left untouched by $U$. Then, if we apply the operator $U$ $t$ times, denoting the resulting unitary by $U_t$, we find
\begin{equation}
    \begin{aligned}
        U_tPU^\dagger_t & =(PX_{n+h}^{-x})^{tx}PZ^{-txz}, \\
        U_tgZ^qU^\dagger_t & = gZ^{(tx+1)q}.
    \end{aligned}
\end{equation}
Notice that the operator $P$ is mapped to a string of operators in which $P$ appears $tx+1$ times, therefore, modulo a phase factor, the final operator is $P^{tx+1}X^{-xtx}Z^{-txz}$. Since $d$ is prime, the equation $tx+1=0\;({\rm mod\;d)}$ always has a solution in $t$. Notice that $U$ only depends on $t$, which in turn depends on $x$, which is determined by $P$. Therefore, by appropriately choosing $t$, the stabilizer generators of the post-measurement state are
\begin{equation}
    \begin{aligned}
        U_t\zeta^{-m}PU^\dagger_t & =\zeta^{s-m}X^xZ^z, \\
        U_tgZ^qU^\dagger_t & = g.
    \end{aligned}
\end{equation}
where $s$ is a suitable phase factor which depends on $x$ and $z$, and we used the fact that $tx=-1\;({\rm mod\;d})$ . Finally, applying a local Clifford unitary on the $n+h$-th qubit we can recover the $Z$ operator. Such a Clifford unitary can be obtained by repeatedly applying the phase gate $S$ to eliminate $Z$, and then applying $F$ to transform the $X$ operators into $Z$. Finally, we apply as many $X$ operators as necessary to obtain the state $\ket{0}$ on the $n+h$-th ancillary qubit. Therefore, we have recovered the initial state and all the operations are Clifford.

\subsection{Proof of lemma \ref{lem:auxlem_2}}
(1. $\iff$ 2.) Since $\Pi_m^P=\frac{1}{d}\sum\zeta^{-mt}P^t$, we have $\bra{\phi_i}\Pi_m^P|\phi_j\rangle=\frac{1}{d}\sum\zeta^{-mt}\bra{\phi_i}P^t|\phi_j\rangle$ for all $i\neq j$, i.e., the coefficients $\bra{\phi_i}\Pi_m^P|\phi_j\rangle$ and $\bra{\phi_i}P^t|\phi_j\rangle$ are related by a discrete Fourier transform, which is invertible, hence $\bra{\phi_i}P^t|\phi_j\rangle$ are vanishing iff $\bra{\phi_i}\Pi_m^P|\phi_j\rangle$ are such. Finally, completeness of $\mcB$ implies that $\langle\phi_i|P|\phi_j\rangle=0$ for all $i\neq j$ iff $\langle\phi_i|P^t|\phi_j\rangle=0$ for all $t$ and $i\neq j$

(2. $\iff$ 3.) Suppose that $P\in \Sb_\mcB$. Then, by definition of $\Sb_\mcB$, $P\in \Sb_{\phi_i}$ for all $i$, that is, $P\ket{\phi_i}=\zeta^c\ket{\phi_i}$ for some $c$. Therefore, $\langle\phi_i|P^t|\phi_j\rangle=\zeta^{c}\braket{\phi_i}{\phi_j}=0$ for all $i\neq j$. Conversely, suppose that $\langle\phi_i|P^t|\phi_j\rangle=0$ for all $i,j$ and $t$ with $P$ a Pauli operator. Then $P$ is diagonal on the basis $\mathcal{B}$ with $P\ket{\phi_i}=\zeta^{c(i)}\ket{\phi_i}$ for some $c(i)$ and all $i$, namely, $P\in\cap_i\Sb_{\phi_i}=\Sb_\mcB$. 

For the last statement, note that item \ref{it:pmstate_ort} is a necessary condition for perfect discrimination. Otherwise we would have two non-orthogonal post-measurement states in the same branch, and the thesis follows by the equivalence 1. $\iff$ 3.

\subsection{Proof of lemma \ref{lem:auxlem3}}


The relation $\dim V_\chi=d^{n-r}$ is a standard fact (see for instance \cite{GHEORGHIU2014505}).
    Now, let $\ket{\phi_b}\in\mcB$ and $Q^{(\ell)}\in\Sb_\mcB$ one of the independent generators. By definition of $\Sb_\mcB$, in particular $Q^{(\ell)}\in\Sb_{\phi_b}$, namely, $\ket{\phi_b}$ is an eigenvector of $Q^{(\ell)}$. This is true for all generators $Q^{(\ell)}$, thus, $\ket{\phi_b}$ is a joint eigenvector of all the generators. This argument holds for all $\ket{\phi_b}\in\mcB$, thus $\ket{\phi_b}\in V_\chi$ for a unique $\chi$, where uniqueness follows by the definition of eigenvector. In particular, this argument implies $\mcB=\cup_\chi(V_\chi\cap\mcB)$. We then have the following chain of equalities, by disjointness of $\mcB\cap V_\chi$
    \begin{equation}
        d^n=|\mcB|=\left|\cup_\chi V_\chi\cap\mcB \right|=\sum_\chi|V_\chi\cap \mcB|.
    \end{equation}
    Note that, since $\dim V_\chi=d^{n-r}$, $\mcB\cap V_\chi$ contains no more than $d^{n-r}$ elements of $\mcB$. Therefore,
     the sum has exactly $d^r$ terms and each of them is at most $d^{n-r}$, none of the terms can be strictly smaller than $d^{n-r}$, thus $|\mcB\cap V_\chi|=d^{n-r}$ and $\mcB\cap V_\chi$ is a basis of $V_\chi$. Finally, the existence of a unitary as described in item \ref{it:auxlem3_2} is also a standard fact in stabilizer code theory (see section 5 of \cite{doi:10.1142/S0129054103002011}).

\subsection{Proof of theorem \ref{thm:criterion}}
    Suppose $\mcB$ is a discriminable stabilizer basis. If by contradiction we would have $\Sb_\mcB=\{\mathds{1}\}$, by \ref{lem:auxlem_2} every non-trivial measurement would cause two post-measurement states to be non-orthogonal. Then it must be $\Sb_\mcB\neq\{\mathds{1}\}$. Now, let us measure all the operators in $\Sb_\mcB$. Since $\mcB$ is perfectly discriminable, $\mcB\cap V_\chi$, which is Clifford-equivalent to $\mcB_\chi\otimes\ket{0}^{\otimes r}$ by lemma \ref{lem:auxlem3}, is perfectly discriminable for all $\chi$, since they are subsets of a perfectly discriminable set. By the ancilla removal lemma \ref{lem:ancilla_removal}, $\mcB_\chi$ is perfectly discriminable.

    On the other hand, Suppose that $\Sb_\mcB\neq \{\mathds{1}\}$ and $\mcB_\chi$ are perfectly discriminable for all $\chi$. First, measure all operators in $\Sb_\mcB$. For every outcome $\chi$, $\mcB_\chi$ is perfectly discriminable, thus, the states in the branch $\mcB\cap V_\chi$ determined by $\chi$ are perfectly discriminable, being Clifford-equivalent to $\mcB_\chi\otimes\ket{0}^{\otimes r}$. Therefore, $\mcB$ is perfectly discriminable.

\section{Proof of theorem \ref{thm:main}}

\subsection{Relation between process functions and unambiguous product bases}

The process functions framework \cite{Baumeler_2016} provides a setting to study the consequences of dropping the assumption of a definite causal order in a classical communication scenario. The correlational scenario is defined as follows: consider $n$ parties, each in his/her own laboratory, who are allowed to perform one round of classical communication by exchanging classical systems with the environment. Each party receives an input $x_k\in\mathcal{X}_k$ from the environment only once and, similarly, feeds one output $a_k\in\mathcal{A}_k$ into it. Moreover, each party has access to an input $i_k\in\mathcal{I}_k$, the local setting, and produces an outcome $o_k\in\mathcal{O}_k$. The basic assumption is that classical probability theory holds locally, which means that the local operations they can apply to their inputs are stochastic processes $p(o_k,a_k|i_k,x_k)$. The environment essentially encodes how the parties communicate with each other, and in the deterministic case it is mathematically modeled by a function $\omega:\bigtimes_k\mathcal{A}_k\rightarrow\bigtimes_k\mathcal{X}_k$. Combining this function with the local interventions of the parties yields the observed correlations
\begin{equation}\label{eq:correlations}
P(\bs{o}|\bs{i})=\sum_{\bs{a},\bs{x}}\delta_{\omega(\bs{a}),\bs{x}}\prod_kp(o_k,a_k|i_k,x_k).    
\end{equation}
A process function is a map $\omega$ that gives a legitimate probability distribution $P(\bs{o}|\bs{i})$, for arbitrary choice of local interventions; otherwise the function is called a quasi-process function. 

Informally speaking, a process function is causal when there exists a party in the global past of all the other parties, and for all such parties, for any of their output, the other parties still communicate in a definite causal order.
More precisely, let $\omega:\bigtimes_k\mathcal{A}_k\rightarrow \bigtimes_k\mathcal{X}_k$ be a process function. Fix an output $b_k$ for a given party $k$ and define $\omega^{b_k}:\mathcal{A}_{\backslash k}\rightarrow \mathcal{X}_{\backslash k}$ as follows
\begin{equation}
    \bs{\omega}^{b_k}(\bs{a}_{\backslash k}):=\bs{\omega}_{\backslash k}(\bs{a}_{\backslash k},f_k(\omega_k(a_{\backslash k}))).
\end{equation}
A crucial point here is that a process function is non-self signaling, which guarantees the well-posedness of the above definition. $\bs{\omega}^{b_k}$ is a legitimate process function called output-reduced process function. Then, the following characterization of causal process functions holds

\begin{theorem}[Theorem 4 \cite{dourdent2025paradoxfreeclassicalnoncausalityunambiguous}]
\label{thm:noncausalPF}
    Let $\bs{\omega}:\mathcal{A}\rightarrow\mathcal{X}$ be a $n$-partite process function. Then, $\bs{\omega}$ is causal if and only if the following conditions are satisfied:
    \begin{enumerate}
        \item \label{it:cond1}there exists a party $k_1$ such that no other party can signal to it, i.e., the input received by the environment is constant
        \begin{equation}
            x_{k_1}=\omega_{k_1}(a_{\backslash k_1})=\overline{x}_{k_1}.
        \end{equation}
        \item \label{it:cond2}For every such party and for every $a_{k_1}\in\mathcal{A}_{k_1}$, the output-reduced process functions $\bs{\omega}^{a_{k_1}]}$ are themselves causal $n-1$-partite process functions.
    \end{enumerate}
\end{theorem}

Given a tensor product of $n$ qudit systems $\mathcal{H}_1\otimes\dots\otimes\mathcal{H}_n$, each of dimension $d_k$, a complete product basis is a set of vectors $\mathcal{S}$ of the form
$\mcB = \{\ket{\phi_i} = \otimes_{k=1}^n\ket{\phi_i^k}\}_{i =1}^{|\mcB|}$ where $|\mcB|=\prod_kd_k$.
For each party $k$, we can group the local vectors into a set $\mathcal{B}^{(k)} = \{\ket{\phi_i^k}\}_{i =1}^{\beta_k}$  of $\beta_k \leq |\mcB|$ distinct elements (so $\mcB^{(k)}$ does not contain repetitions of the same element). This set can be further divided into $|\mathcal{X}_k|$ subsets $\mathcal{B}^{(k)}_{x_k}$ such that $\mcB^{(k)} = \cup_{x_k = 0}^{|\mathcal{X}_k| - 1}\mcB^{(k)}_{x_k}$
and all vectors in a given subset are mutually orthogonal. Since $\mathcal{B}$ is a basis, each subset 
$\mcB^{(k)}_{x_k}$ contains exactly $d_k$ elements (the local dimension) and is itself a local basis. With this notation, the definition of a unambiguous product basis is given the following.
\begin{definition}
An unambiguous basis is a product basis such that, for every party $k$, two local vectors are orthogonal if and only if they belong to the same setting, i.e., $\langle\phi^k_{j'}|\phi^k_j\rangle=0$ iff $x_k^{j'}=x_k^j$.    
\end{definition}
Stabilizer product basis are such that local bases are mutually unbiased. This means that two vectors $\ket{\phi_1}$, $\ket{\phi_2}$ belonging to two different stabilizer basis satisfy $|\braket{\phi_1}{\phi_2}|^2=\frac{1}{d}$. This implies that the above definition, therefore SPBs are unambiguous. This finally implies that for every SPB $\mcB$ there exist a process function $\omega$ and a set of local Clifford unitaries $\{U_k^{x_k}\}_{k=1}^n$ as follows \cite{dourdent2025paradoxfreeclassicalnoncausalityunambiguous}
\begin{equation}\label{eq:pf_rep_dup}
\mcB=\{\bigotimes_kU_k^{\omega_k(\boldsymbol a_{\backslash k})}\ket{\boldsymbol a}|\bs{a}\in\bs{\mathcal{A}}\}.
\end{equation}
where $U_k^{x_k}\in \langle F,S\rangle$, with $F$ and $S$ the generalized Hadamard and phase gates \cite{doi:10.1142/S0129054103002011,Clark_2006}

\subsection{Proof of theorem \ref{thm:main}}

    If $\omega$ is causal, it can be easily seen that $\mcB$ is perfectly discriminable with a 1-way LOCC protocol. To show the other implication we proceed by induction on $n$. For $n=2$ the statement is trivially true, because all 2-partite process functions are causal \cite{Baumeler_2016}. Now, let us suppose that the statement is true for $n-1$. The inductive hypothesis is that whenever a stabilizer product basis on $n-1$ qudits is discriminable, the corresponding $(n-1)$-process function is causal. Now,
    by hypothesis $\Sb_\mcB\neq\{\mathds{1}\}$ and all its generators can be taken to be local, because $\mcB$ is a product basis.
   Indeed, let $Q\in\Sb_\mcB$ be a non trivial Pauli operator. Each $\Sb_{\phi_i}$ contains a local Pauli operator on the non-trivial factor $\ell$, say $Q_\ell^{(i)}$. Such operator must commute with $Q$, because both belong to the same $\Sb_{\phi_i}$, and this holds for all $i$. For this reason, for each $\Sb_{\phi_i}$, $Q_\ell^{(i)}$ must be a power of the same local operator $Q_\ell$. Therefore, $Q_\ell$ belongs to $\Sb_{\phi_i}$ for all $i$.
    Now, consider a non-trivial generator $Q_\ell$ for some qubit labeled by $\ell$ such that $Q_\ell\in\Sb_\phi$ for all $\phi\in\mcB$. 
    Then, the local basis of $\ell$ is the same for all states $\ket{\phi}$ in $\mcB$, implying that $\omega_{\ell}$ is constant, i.e., that $\omega$ has a global past. Thus condition \ref{it:cond1} of theorem \ref{thm:noncausalPF} is met.
    For every branch determined by the outcome $m$ the discrimination problem is reduced to the discrimination of the set $\mcB_m\otimes\ket{\phi_m}$, where $\ket{\phi_m}$ is an eigenvector of $Q_\ell$, i.e., to perfect discrimination of $\mcB_m$. Now, $\ket{\phi_m}=U_\ell^{\omega_\ell}\ket{b_\ell}$ for some fixed $b_\ell$, one for each different outcome $m$. Thus, $\mcB_m$ is determined by the $(n-1)$-partite output-reduced process function obtained from $\omega$, where the output of the party $\ell$ is $b_\ell$. By hypothesis, each $\mcB_m$ is perfectly distinguishable, and by the inductive hypothesis the $(n-1)$-partite process $\omega^{b_\ell}$ is casual, and this holds for every $b_\ell$. 
    Also, observe that a component $\omega_\ell$ is constant if and only if a local Pauli $Q_\ell$ belongs to $\Sb_\mcB$. Thus, the above argument covers all parties with constant component, and condition \ref{it:cond2} of theorem \ref{thm:noncausalPF} is also satisfied. We conclude that $\omega$ is a causal $n$-partite process.

\end{document}